\documentclass[referee]{aastex}
\usepackage{natbib}
\usepackage{amsmath}
\received{}
\revised{}
\accepted{}
\submitjournal{ApJ}

\shorttitle{Reactive Desorption of CO Hydrogenation Products}
\shortauthors{Chuang et al.}

\begin{document}

\title{Reactive Desorption of CO Hydrogenation Products under Cold Pre-stellar Core Conditions}

\correspondingauthor{K.-J. Chuang}
\email{chuang@strw.leidenuniv.nl}

\author{K.-J. Chuang}
\affil{Sackler Laboratory for Astrophysics, Leiden Observatory, Leiden University, P.O. Box 9513, NL-2300 RA Leiden, the Netherlands}
\affiliation{Leiden Observatory, Leiden University, P.O. Box 9513, NL-2300 RA Leiden, the Netherlands}
\author{G. Fedoseev}
\affiliation{INAF – Osservatorio Astrofisico di Catania, via Santa Sofia 78, 95123 Catania, Italy}

\author{D. Qasim}
\affiliation{Sackler Laboratory for Astrophysics, Leiden Observatory, Leiden University, P.O. Box 9513, NL-2300 RA Leiden, the Netherlands}

\author{S. Ioppolo}
\affiliation{School of Electronic Engineering and Computer Science, Queen Mary University of London, Mile End Road, London E1 4NS, UK}
\affiliation{School of Physical Sciences, The Open University, Walton Hall, Milton Keynes MK7 6AA, UK}

\author{E.F. van Dishoeck}
\affiliation{Leiden Observatory, Leiden University, P.O. Box 9513, NL-2300 RA Leiden, the Netherlands}

\author{H. Linnartz}
\affiliation{Sackler Laboratory for Astrophysics, Leiden Observatory, Leiden University, P.O. Box 9513, NL-2300 RA Leiden, the Netherlands}



\begin{abstract}
{}The astronomical gas-phase detection of simple species and small organic molecules in cold pre-stellar cores, with abundances as high as $\sim$$10^{-8}-10^{-9}$ n$_\text{H}$, contradicts the generally accepted idea that at $10$ K, such species should be fully frozen out on grain surfaces. A physical or chemical mechanism that results in a net transfer from solid-state species into the gas phase offers a possible explanation. Reactive desorption, i.e., desorption following the exothermic formation of a species, is one of the options that has been proposed. In astronomical models, the fraction of molecules desorbed through this process is handled as a free parameter, as experimental studies quantifying the impact of exothermicity on desorption efficiencies are largely lacking. In this work, we present a detailed laboratory study with the goal of deriving an upper limit for the reactive desorption efficiency of species	involved in the CO-H$_2$CO-CH$_3$OH solid-state hydrogenation reaction chain. The limit for the overall reactive desorption fraction is derived by precisely investigating the solid-state elemental carbon budget, using reflection absorption infrared spectroscopy and the calibrated solid-state band-strength values for CO, H$_2$CO and CH$_3$OH. We find that for temperatures in the range of $10$ to $14$ K, an upper limit of $0.24\pm 0.02$ for the overall elemental carbon loss upon CO conversion into CH$_3$OH. This corresponds with an effective reaction desorption fraction of $\leq$$0.07$ per hydrogenation step, or $\leq$$0.02$ per H-atom induced reaction, assuming that H-atom addition and abstraction reactions equally contribute to the overall reactive desorption fraction along the hydrogenation sequence. The astronomical relevance of this finding is discussed.
\end{abstract}



\section{Introduction}\label{ch6Introduction}

Surface chemistry in the earliest phase of molecular cloud evolution is determined by a rather low particle density ($\sim$$10^3-10^4$ cm$^{-3}$) at temperatures around $15$ K. In such environments, chemical processes on dust grains are largely dominated by H-atom addition reactions, forming an H$_2$O-rich polar ice layer on silicate dust grains by recombining oxygen and hydrogen atoms \citep{Tielens1982, Tielens1991, Ioppolo2008, Ioppolo2010, Miyauchi2008, Cuppen2010, Oberg2011a, vanDishoeck2013, Boogert2015, Linnartz2015}. During this stage, other molecules, like NH$_3$ and CH4, also form in N+H and C+H addition reactions \citep{Hasegawa1992, Hiraoka1995, Hiraoka1998, Hidaka2011, Fedoseev2015a}. In the next phase, temperatures become as low as $\sim$$10$ K and densities increase ($\sim$$10^{4-5}$ cm$^{-3}$); gaseous CO accretes onto the pre-formed H$_2$O-rich ice and forms a CO-rich apolar ice coating \citep{Pontoppidan2006}, an event often referred to as the \textquoteleft catastrophic CO freeze-out stage.\textquoteright~The timescale for the CO freeze-out phase is about $10$$^{4-5}$ years ($1$$\times$$10^9$/n$_\text{H}$ years where \textit{n}$_\text{H}$ = $2$\textit{n}(H$_2$) + \textit{n}(H), \citealt{Willacy1998}), which is shorter than the typical lifetime of a low-mass prestellar core ($\sim$$4.5$$\times$$10^5$ years, \citealt{Enoch2008}). The CO coating is regarded as a starting point in the hydrogenation scheme that leads primarily to the formation of H$_2$CO and CH$_3$OH through successive H-atom additions. This reaction scheme has been studied in much detail before, in a gas-grain model by \citet{Tielens1982}, in astrochemical simulations and theoretical studies \citep{Charnley1997, Cuppen2009, Chang2012}) and in a number of systematic laboratory experiments \citep{Hiraoka1994, Watanabe2002, Zhitnikov2002, Fuchs2009}. The solid-state laboratory and astrochemical modeling conclusions are in agreement with the observed CH$_3$OH ice and gas abundances in star-forming regions that cannot be explained through a pure gas-phase radiative association route \citep{Garrod2006, Geppert2006}. Analysis of Very Large Telescope ice data shows that CH$_3$OH and CO must be mixed in interstellar ices and exist in a water poor environment \citep{Cuppen2011, Penteado2015}, fully consistent with the picture that methanol is formed via sequential hydrogenation of accreted CO molecules. Moreover, recent theoretical and laboratory studies revealed that recombination of the reactive intermediates, HCO, CH$_3$O, and CH$_2$OH formed through H-atom addition and abstraction reactions along the CO-H$_2$CO-CH$_3$OH chain, also result in the formation of larger complex organic molecules (COMs), like ethylene glycol, glycolaldehyde, methyl formate, and even glycerol in the solid state \citep{Garrod2006, Woods2012, Butscher2015, Fedoseev2015b, Fedoseev2017, Chuang2016, Chuang2017}.

For the typical pressure and temperature conditions in dark and dense clouds, all of these species except H$_2$ and He are expected to be fully depleted into the solid state. However, astronomical observations show unexpectedly large abundances of gaseous CH$_3$OH and COMs \citep{Oberg2010, Bacmann2012, Cernicharo2012, Jimenez-Serra2016}. In dark and pre-stellar clouds, it is too cold for thermal desorption to take place; thus, non-thermal mechanisms need to be considered to explain the observed gas-phase abundances, but effective low-temperature desorption processes remain poorly understood. 

Different mechanisms have been proposed, investigated, and discussed, to explain how frozen species can sublimate. Until recently, UV induced photodesorption was considered to be an efficient process for both non-dissociative desorption, i.e., CO photodesorption, and dissociative processes eventually followed by fragment recombination in the ice or kick-out processes, i.e., H$_2$O photodesorption \citep{Andersson2008, Arasa2015}. However, experiments on methanol ice exhibit substantial molecular fragmentation, including CH$_3$O, CH$_2$OH, and CH$_3$, and with photodesorption rates far too low to transfer substantial amounts of intact CH$_3$OH into the gas phase \citep{Bertin2016, Cruz-Diaz2016}. It is expected that this will also apply to COMs larger than methanol; photodesorption and photodissociation are here clearly connected. A possible way to circumvent fragmentation upon photo-excitation is UV induced codesorption; the excitation of one species results in desorption of another species. For mixed CO and N$_2$ ices, N$_2$ photocodesorption rate (defined as indirect DIET photodesorption rate) was found to be a highly effective process, i.e., $\sim$$10^{-2}$ molecule cm$^{-2}$ photon$^{-1}$ for photon energies between $7.9$ and $9.5$ eV \citep{Bertin2013}, but in ice mixtures of CO and CH$_3$OH no methanol could be detected upon CO excitation \citep{Bertin2016}. In a similar way, low-temperature thermal codesorption of a less volatile species like methanol with CO was only found to yield very low upper limits of $7$$\times$$10^{-7}$ molecules per CO \citep{Ligterink2018}. Local heating of an ice mantle induced by the impacting cosmic rays offers another possible mechanism \citep{Leger1985, Ivlev2015}, but given the high energies involved, it is rather unlikely that this would be a less destructive process than photodesorption, and this process is generally less effective for strongly bound molecules like CH$_3$OH than for CO. An interesting alternative is reactive desorption.

Reactive desorption may contribute to the gas-phase enrichment of both simple and complex species; directly, upon formation, formaldehyde, methanol or other COMs are instantly released into the gas phase because of the available excess energy of several eV, or indirectly, by releasing intermediates that cannot be efficiently formed in gas-phase chemical networks, but may act as precursor in COM gas-phase networks. A reactive codesorption process cannot be excluded either, but the excess energy upon bond formation may rapidly dissipate into the ice lattice before being absorbed by another species. These three scenarios may take place simultaneously, but with different efficiencies.

In astrochemical models, the reactive desorption fraction, i.e., the efficiency of product desorption from the surface after reaction, is usually a free parameter, chosen in a range of $0.01-0.1$ to account for the steady-state gas-phase abundances of simple species as observed in the later stages of dense clouds \citep{Garrod2007, Vasyunin2013b}. The initial focus was on energy release upon H$_2$ formation \citep{Duley1993}, and later extended to other reactions \citep{Garrod2007, Vasyunin2013b, Cazaux2015, Minissale2016b, Fredon2017}. In the latter work, reactive desorption of species was also proposed as a starting point for low-temperature gas-phase formation of COMs (see also \citealt{Balucani2015, Taquet2016, Rivilla2017, Vasyunin2017}). Currently, the relative solid-state and gas-phase formation efficiencies for dark cloud conditions are under renewed debate.

\begin{figure}
	\begin{center}
		\includegraphics[width=120mm]{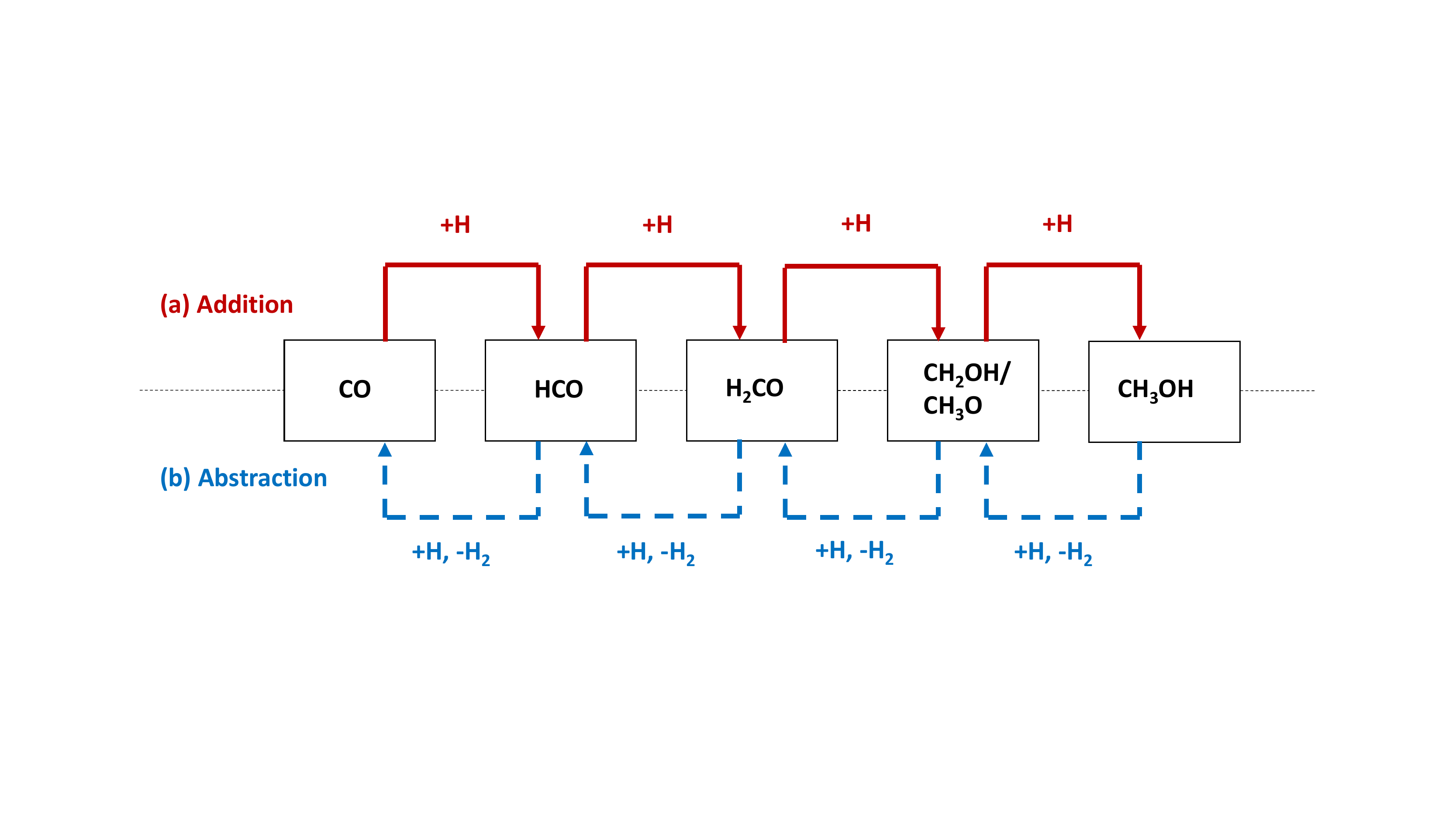}
		\vspace{-1.5cm}
		\caption{
			CO-H$_2$CO-CH$_3$OH hydrogenation sequence scheme. The solid line and dashed line are for H-atom (a) addition and (b) abstraction reactions, respectively.
		}
		\label{Fig6.1}
	\end{center}
\end{figure}

So far, there have been few attempts to experimentally study the reactive desorption of the two main CO hydrogenation products, H$_2$CO and CH$_3$OH. There is a reason for this: the simultaneous occurrence of H-atom induced addition and abstraction reactions, i.e., forward and backward steps along the reaction chain (Figure \ref{Fig6.1}), make it hard to de-convolute the individual reaction steps.

Previously reported experimental desorption fractions for reactive steps in the CO hydrogenation chain are high. \citet{Hidaka2004} stated that the H$_2$CO reactive desorption fraction during hydrogenation could be as high as $60$\% to rationalize their finding that the column density of the newly formed CH$_3$OH product was less than the consumed column density of the pre-deposited H$_2$CO, i.e., $\Delta \textit{N}$ (CH$_3$OH) / $\Delta \textit{N}$ (H$_2$CO) $<$ $1$. \citet{Minissale2016b} found a reactive desorption fraction of $40$\% for the single reaction HCO + H $\longrightarrow$ CO + H$_2$. This reaction was studied in the sub-monolayer region on a bar of oxidized highly oriented pyrolytic graphite substrate using quadrupole mass spectrometry. For a more astronomically relevant thick ice (several tens of monolayers) these conditions are likely not fully representative and therefore reactive desorption fractions may be substantially different as well.

The work presented here is motivated by the lack of experimentally determined reactive desorption fractions for hydrogenation reactions in multi-layered CO-rich interstellar ice analogs, i.e., for an astrochemically relevant ice thickness of $\sim$$0.01$ $\mu$m \citep{Boogert2015}. Rather than trying to disentangle the individual reactions and to measure the individual reaction products in the gas phase, we focus on the full CO-H$_2$CO-CH$_3$OH hydrogenation sequence. This is realized by estimating as precisely as possible, the difference in solid-state elemental carbon budget between the amount available from the original species and the resulting solid-state hydrogenation products. We assume that the observed difference is fully or at least partially due to reactive desorption, i.e., the resulting value reflects an upper limit. To perform such experiments, it is necessary to derive setup specific band-strength values in cm molecule$^{-1}$ for each of the involved species and their ice mixtures.

In the following section the experimental procedure is described. The results are presented in Section \ref{ch6Results and Discussion}. In the final section the astronomical relevance of the data and conclusions are presented.

\section{Experimental}\label{ch6Experimental}
\subsection{Approach}\label{ch6Approach}

The total consumption of elemental carbon of the original species should be equal to the total amount of elemental carbon of the formed species, both in the solid state and, after reactive desorption, in the gas phase:
\begin{equation}\label{Eq6.1}
\textit{N}_{\text{consumed}}\text{(C)}=\Delta\textit{N}_{\text{solid}}\text{(CO)}=\Delta\textit{N}_{\text{solid}} \rm{(H_{2}CO)}+\Delta\textit{N}_{\text{solid}}\rm(CH_3OH)+\textit{N}_{RD}\text{(C)}
\end{equation}
where $\Delta$$\textit{N}_\text{solid}$(CO) is the column density of CO consumption, $\Delta$$\textit{N}_\text{solid}$(H$_2$CO)($\Delta$$\textit{N}_\text{solid}$(CH$_3$OH)) is the net column density of the H$_2$CO (CH$_3$OH) formation, and $\textit{N}_{RD}$(C) is the column density of carbon-bearing species upon reactive desorption mechanism after H-atom bombardment. This means that an upper limit for $\textit{N}_{RD}$(C)/$\textit{N}_\text{consumed}$(C) can be derived by comparing accurately the total difference in CO abundance and H$_2$CO and CH$_3$OH abundances. In practice, Equation (\ref{Eq6.1}) is only an approximation, as it does not take into account other loss channels, such as the formation of other products than H$_2$CO and CH$_3$OH; reactive intermediates can recombine and form COMs larger than methanol, as recently shown by \citet{Chuang2016} and \citet{Fedoseev2017}. The impact of this restriction is discussed in Section \ref{ch6Results and Discussion}.

\subsection{Experimental Setup}\label{ch6Experimental Setup}

All experiments are performed using SURFRESIDE$^2$, an ultra-high vacuum setup fully optimized to study non-energetic atom addition reactions in interstellar ice analogs at cryogenic temperatures. Details of the design and operation of SURFRESIDE$^2$ are available from \citet{Ioppolo2013} and \citet{Fedoseev2016}. The base pressure in the main chamber is $\sim$$10^{-10}$ mbar, and the corresponding H$_2$O contamination through background gas accretion is negligible ($<$$6.3$$\times$$10^{10}$ molecule cm$^{-2}$ s$^{-1}$). A gold-plated copper substrate is mounted in the center of the main chamber and cooled by a recently upgraded closed-cycle helium cryostat that allows to vary the substrate temperature between $8$ and $450$ K. A silicon diode is used to monitor the temperature with $0.5$ K absolute accuracy. It should be noted that the temperatures used in the present manuscript are about $\sim$$2$ K lower than the corresponding numbers used in \citet{Fuchs2009} new and more precise thermal sensors were installed on the substrate. Gaseous species, i.e., H$_2$CO obtained by thermal decomposition of paraformaldehyde (Sigma-Aldrich, $95$\%) at $60$ $^\circ$C$-$$80$ $^\circ$C under vacuum, CH$_3$OH purified through multiple freeze-pump-thaw cycles and CO (Linde 2.0), are admitted into the UHV chamber through high-precision leak valves mounted under an angle of $22$$^\circ$ from the substrate surface normal. The ices are grown with sub-monolayer precision with CO, H$_2$CO and CH$_3$OH deposition rates of $\sim$$1.7$$\times$$10^{12}$, $\sim$$1.4$$\times$$10^{12}$ and $\sim$$2.2$$\times$$10^{12}$ molecules cm$^{-2}$ s$^{-1}$, respectively. The corresponding ice column densities for CO, H$_2$CO, and CH$_3$OH are $\sim$$6.0$$\times$$10^{15}$, $\sim$$5.0$$\times$$10^{15}$, and $\sim$$8.0$$\times$$10^{15}$ molecules cm$^{-2}$, respectively. The details of the exact deposition rate calculations are provided later.

After a preset ice thickness of the precursor species is realized, H-atoms are introduced using a Hydrogen Atom Beam Source \citep{Tschersich2000}, mounted in a second UHV chamber. A nose-shape quartz pipe is placed along the path to efficiently quench and thermalize excited H-atoms and nondissociated molecules through multiple collisions with the walls of the pipe. The used H-atom flux in this work amounts to $\sim$$8$$\times$$10^{12}$ atoms cm$^{-2}$ s$^{-1}$ with an absolute uncertainty of $<$$50$\% \citep{Ioppolo2013} and H-atoms reach the surface under an incident angle of 45$^\circ$ from the substrate surface normal.

The ice composition is monitored in situ, before and during H-atom bombardment by Fourier Transform Reflection-Absorption InfraRed Spectroscopy in the range from $700$ to $4000$ cm$^{-1}$, with $1$ cm$^{-1}$ resolution. The ice column density $\textit{N}$ in cm$^{-2}$ is derived from a modified Beer's law:
\begin{equation}\label{Eq6.2}
\textit{N}= \frac{\log 10 \cdot \int \textit{Abs}(\nu) d\nu}{\textit{A'}}
\end{equation}
where $\textit{Abs($\nu$)}$ is the band absorbance, $\textit{d$\nu$}$ is the wavenumber differential in cm$^{-1}$, and $\textit{A'}$ is the calibrated band strength in cm molecule$^{-1}$ in reflection mode. This RAIR band strength cannot be taken from the literature data available for the IR transmission spectroscopy, as the signals obtained in reflection are enhanced through substrate dipole couplings. Moreover, the determination of RAIR band strengths from transmission values cannot be realized by only compensating for different effective IR pathways in the ice. Therefore, a series of extra experiments (see Section \ref{ch6Band Strength Measurement for Pure Ices}) has been performed to determine the band strength for each of the involved species for our specific experimental settings by using the interference pattern of a HeNe laser that is reflected from the growing ice sample at $10$ K \citep{Baratta1998, Brunetto2008, Fulvio2009, Bouilloud2015}. For pure ice the absolute column density is calculated by the equation:
\begin{equation}\label{Eq6.3}
\textit{N}= \frac{\textit{d} \cdot \textit{$\rho$} \cdot \textit{N}_a}{\textit{M}}
\end{equation}
where $\textit{d}$ is the thickness of ice in nm, \textit{$\rho$} is the density in g cm$^{-3}$, $\textit{N$_a$}$ is the Avogadro's constant ($6.022$$\times$$10^{23}$ mol$^{-1}$), and $\textit{M}$ is the molar mass of the species. The ice thickness ($D_{\text{growing}}$) is experimentally determined by laser refractive interference,
\begin{equation}\label{Eq6.4}
\textit{N}= \textit{k} \ \frac{\textit{$\lambda$}}{2 \textit{n} \cdot \cos (\textit{$\theta$}_f) }
\end{equation}
where $\textit{$\lambda$}$ = $632.8$ nm is the HeNe laser wavelength, $\textit{n}$ is the refractive index of a specific ice, $\textit{$\theta_f$}$ $\cong$ $3$ is the angle of refraction in the ice in degrees, and $\textit{k}$ is the number of involved fringes \citep{Hollenberg1961, Westley1998}. The exact band-strength values may vary with changing ice composition, structure, and temperature. The reported absolute uncertainty of this method is $5$\%, a value that is largely determined by the density and refractive index uncertainty \citep{Baratta1998, Fulvio2009}. Moreover, as the same technique is used to determine CO, H$_2$CO and CH$_3$OH band strengths, the relative error between values is expected to be lower. This makes it possible, in principle, to establish the obtained $\textit{N}_{RD}$(C)/$\textit{N}_\text{consumed}$(C) with high confidence, even when chemistry changes relative ice compositions.

\subsection{Band strength measurement for pure ices}\label{ch6Band Strength Measurement for Pure Ices}

\begin{figure}
	\begin{center}
		\includegraphics[width=120mm]{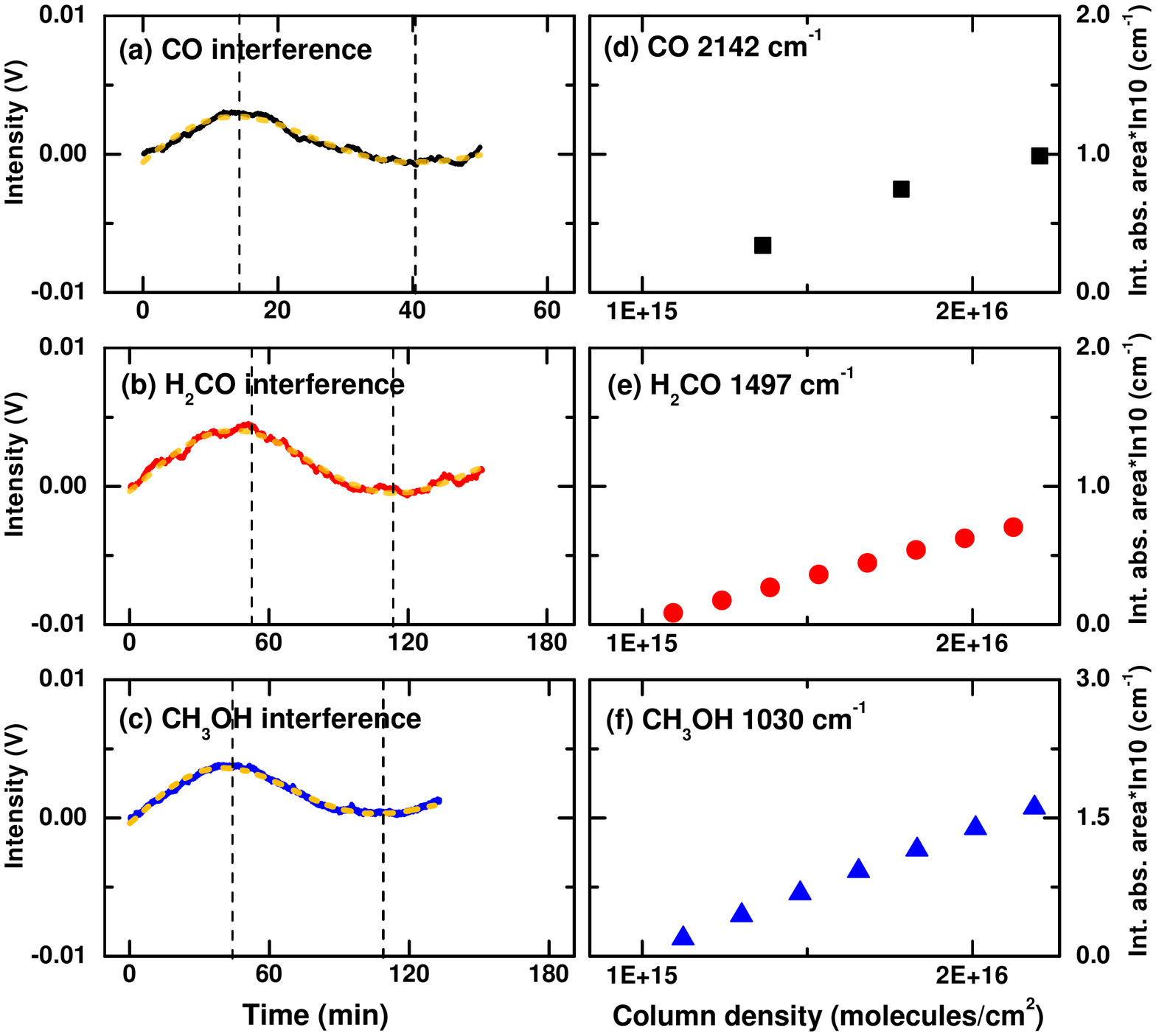}
		\caption{
			Left: typical examples of the obtained interference fringes as function of time and their corresponding sinusoidal fittings (yellow dashed line) during $10$ K substrate deposition of pure (a) CO, (b) H$_2$CO, and (c) CH$_3$OH. The vertical dashed lines (black) indicate the time for maximum and minimum amplitude. Right: the integrated absorption area for deposited (d) CO ($2142$ cm$^{-1}$ band), (e) H$_2$CO ($1497$ cm$^{-1}$ band) and (f) CH$_3$OH ($1030$ cm$^{-1}$ band) as a function of column density for the first $7$ minutes of deposition.
		}
		\label{Fig6.2}
	\end{center}
\end{figure}

The left-hand panel of Figures \ref{Fig6.2}(a)-(c) shows the HeNe laser interference pattern that chang-es with time as the ice thickness increases upon deposition of CO, H$_2$CO, and CH$_3$OH onto the $10$ K substrate. One single fringe is typically observed for the low-deposition rates used in this study. Ice density (\textit{$\rho$}) and refractive index ($\textit{n}$) may vary for different conditions \citep{Bossa2015}. It is important to note that the precise ice density in the amorphous phase is generally unknown for the species studied in this work. Here, the commonly accepted values in literature are taken as good approximate values. For CO, the ice density reported by \citep{Roux1980} at $20$ K is used. Their deposition conditions may have been different from those in our work. For H$_2$CO, no density value has been reported in the solid phase, thus the value measured in liquid phase is employed \citep{Weast1985}. For CH$_3$OH, the ice density is derived from the value available for its crystalline structure (Mate2009). Based on the density and refractive index values summarized in \citep{Bouilloud2015} $0.80$$\pm$$0.01$ g cm$^{-3}$ and $1.25$$\pm$$0.03$ for CO, $0.81$$\pm$$0.03$ g cm$^{-3}$ and $1.33$$\pm$$0.04$ for H$_2$CO, and $1.01$$\pm$$0.03$ g cm$^{-3}$ and $1.33$$\pm$$0.04$ for CH$_3$OH \textemdash~CO, H$_2$CO, and CH$_3$OH deposition rates are derived with values of $\sim$($1.3$$\pm$$0.1$)$\times$$10^{14}$, $\sim$($4.7$$\pm$$0.4$)$\times$$10^{13}$, and $\sim$($5.6$$\pm$$0.5$)$\times$$10^{13}$ molecules cm$^{-2}$ s$^{-1}$. Note that these values differ in these experiments from the regular hydrogenation settings. The choice of the applied density and refractive index values is not ideal, but given the lack of more precise values, this is the best we can provide. If better values become available, the final ratio of $\textit{N}_{RD}$(C)/$\textit{N}_\text{consumed}$(C) can be easily re-evaluated.

The corresponding CO, H$_2$CO, and CH$_3$OH plots, showing the column density upon ice deposition (horizontal axis) and the increasing IR absorption area (vertical axis) over time are presented in the right-hand panels of Figures \ref{Fig6.2}(d)-(e). As explained in Section \ref{ch6Experimental} (Equation \ref{Eq6.2}), from these plots the A', band-strength value, can be derived from molecule specific RAIR bands, following a linear fit. The fitted slopes for (d) CO ($2142$ cm$^{-1}$), (e) H$_2$CO ($1497$ cm$^{-1}$), and (f) CH$_3$OH ($1030$ cm$^{-1}$) yield absorption strength values A' of ($5.2$$\pm$$0.3$)$\times$$10^{-17}$, ($3.2$$\pm$$0.3$)$\times$$10^{-17}$, and ($7.1$$\pm$$0.6$)$\times$$10^{-17}$ cm molecule$^{-1}$, respectively. It should be noted that these values are only safe to use for identical experimental circumstances, i.e., identical IR incident angles, surface materials, etc.

\citet{Kerkhof1999} showed that the band strength of methanol in an ice mixture of CH$_3$O-H:CO2($1$:$1$) and in pure CH$_3$OH ice does not vary. We checked this by performing a set of control experiments in which the CO, H$_2$CO, and CH$_3$OH band-strength values are compared with those derived in CO:H$_2$CO ($1$:$1$), CO:CH$_3$OH ($1$:$1$), and CO:H$_2$CO:CH$_3$OH ($1$:$1$:$1$) mixed ices. This is realized in the following way. At first, a constant deposition rate of CO molecules is used through one of the dosing lines. Then, sequentially, H$_2$CO and CH$_3$OH are introduced through other dosing lines growing an ice mixture during co-deposition. Each time a new species is allowed into the main chamber, and any changes in the CO absorbance integral area are carefully monitored. Then, in a similar way, changes in the H$_2$CO and CH$_3$OH absorbance integral area are measured. These changes are linearly correlated with their increasing column densities. From this, it is found that the CO, H$_2$CO, and CH$_3$OH band strengths are rather constant, with variations that do not exceed $2$\% in CO, CO:H$_2$CO ($1$:$1$), CO:CH$_3$OH ($1$:$1$), and CO:H$_2$CO:CH$_3$OH ($1$:$1$:$1$) ices. RAIR spectra are used only as long as signals are far from saturation, to guarantee a linear correlation with the corresponding column densities \citep{Teolis2007}.

\section{Results and discussion}\label{ch6Results and Discussion}
\subsection{Elemental carbon budget in CO hydrogenation experiments}\label{ch6Elemental Carbon Budget in CO Hydrogenation Experiments}

\begin{figure}
	\begin{center}
		\includegraphics[width=98mm]{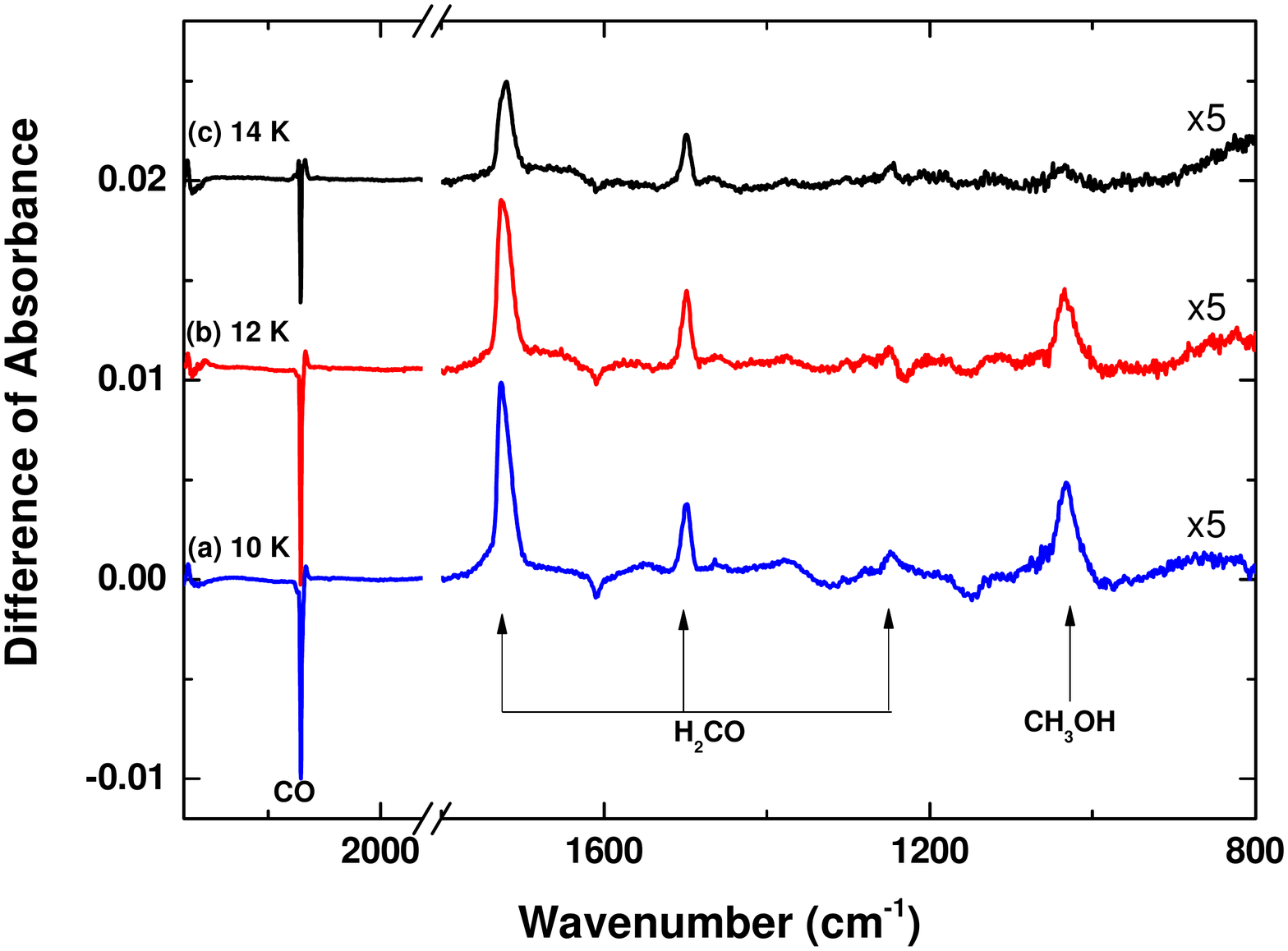}
		\caption{
			RAIR difference spectra obtained after the exposure of $6$$\times$$10^{15}$molecules cm$^{-2}$ pre-deposited CO to an H-atom flux of $\sim$$8$$\times$$10^{12}$ atoms cm$^{-2}$ s$^{-1}$ for $180$ minutes at (a) $10$ K, (b) $12$ K, and (c) $14$ K. Negative and positive signals reflect net consumption and formation, respectively.
		}
		\label{Fig6.3}
	\end{center}
\end{figure}

Figure \ref{Fig6.3} presents RAIR difference spectra obtained after hydrogenation of pre-deposited CO ice for $180$ minutes by an H-atom flux of $\sim$$8$$\times$$10^{12}$ atoms cm$^{-2}$ s$^{-1}$ at (a) $10$ K, (b) $12$ K, and (c) $14$ K, respectively. The negative peak at $2142$ cm$^{-1}$ shows the consumption of the originally predeposited CO ice, and the positive peaks indicate the formation of two newly formed main products, i.e., H$_2$CO ($1727$, $1497$, and $1250$ cm$^{-1}$) and CH$_3$OH ($1030$ cm$^{-1}$). Very similar results have been reported and discussed extensively in previous studies of \citet{Watanabe2002} and \citet{Fuchs2009}. No entrapment of the formed intermediate radicals such as HCO, CH$_3$O, and CH$_2$OH is found, as these easily react with free H-atoms or other radical species.

\begin{figure}
	\begin{center}
		\includegraphics[width=98mm]{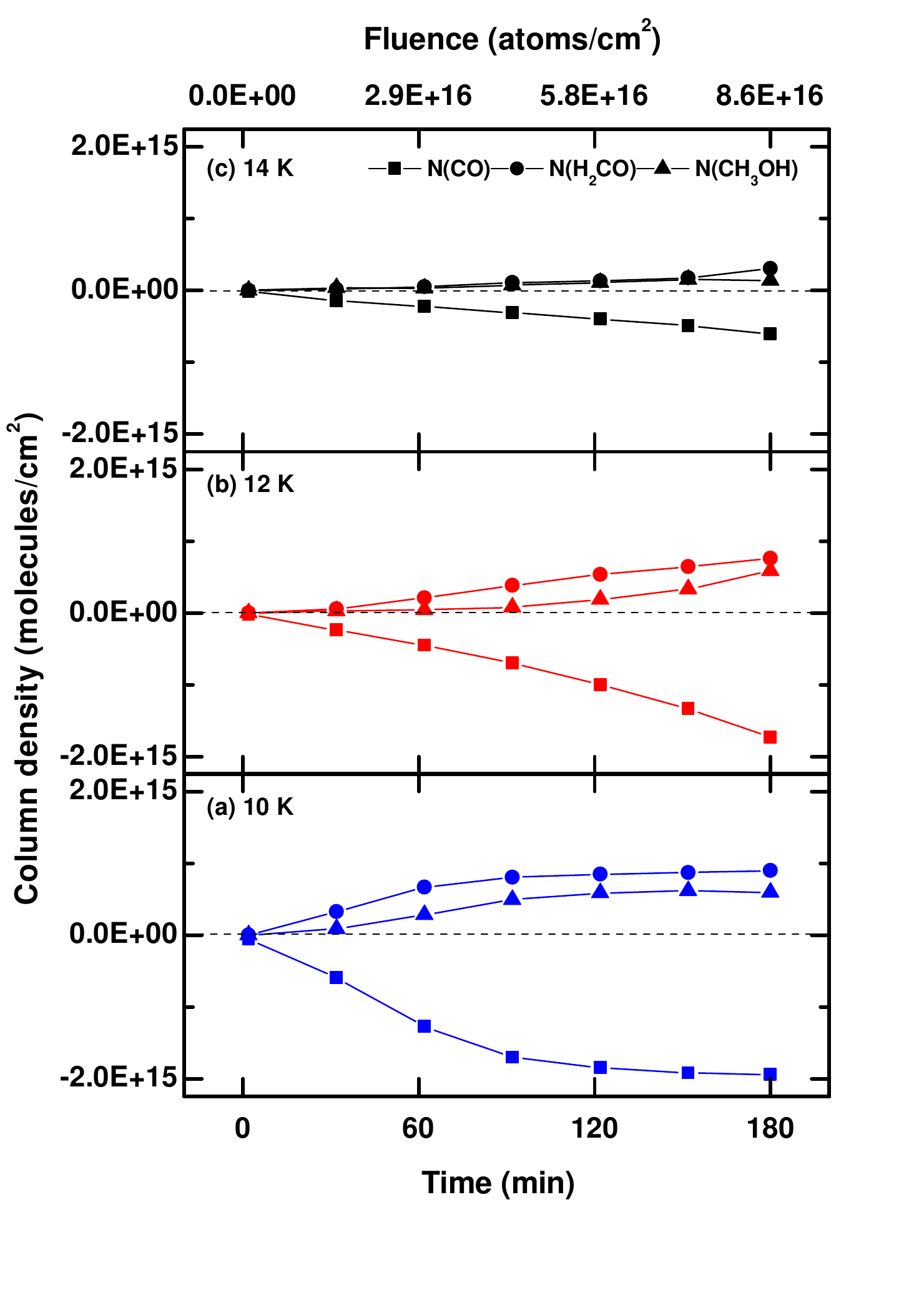}
		\caption{
			Evolution of the surface column density of CO, H$_2$CO, and CH$_3$OH during a CO hydrogenation experiment with an H-atom flux of $\sim$$8$$\times$$10^{12}$ atoms cm$^{-2}$ s$^{-1}$ over $180$ minutes at (a) $10$ K, (b) $12$ K, and (c) $14$ K, respectively.	
		}
		\label{Fig6.4}
	\end{center}
\end{figure}

Figure \ref{Fig6.4} shows the changing column density obtained from the integrated RAIR absorption areas of the CO ($2142$ cm$^{-1}$), H$_2$CO ($1497$ cm$^{-1}$), and CH$_3$OH ($1030$ cm$^{-1}$) signals during H-atom accretion. The negative CO consumption peak at $10$ K clearly saturates due to the limited penetration depth of H-atoms impacting on the surface. This observation is in agreement with previous findings \citep{Fuchs2009}, indicating that the CO consumption rate exhibits a strong temperature dependence; a lower temperature results in a higher CO conversion rate, fully in line with the proposed CO-H$_2$CO-CH$_3$OH reaction findings presented in the literature \citep{Watanabe2004, Fuchs2009}. This temperature dependence is explained by the different life (i.e., residence) time of H-atoms on the ice surface before desorbing from the ice surface or recombining to H$_2$ through interactions with other H-atoms. At $10$ K and after $120$ minutes of H-atom exposure, the CO depletion behavior starts slowing down, reaching a maximum value at approximately $2$$\times$$10^{15}$ molecules cm$^{-2}$ ($180$ minutes), reflecting that only the few upper layers of CO molecules are accessible to accreting H-atoms.

\begin{figure}
	\begin{center}
		\includegraphics[width=98mm]{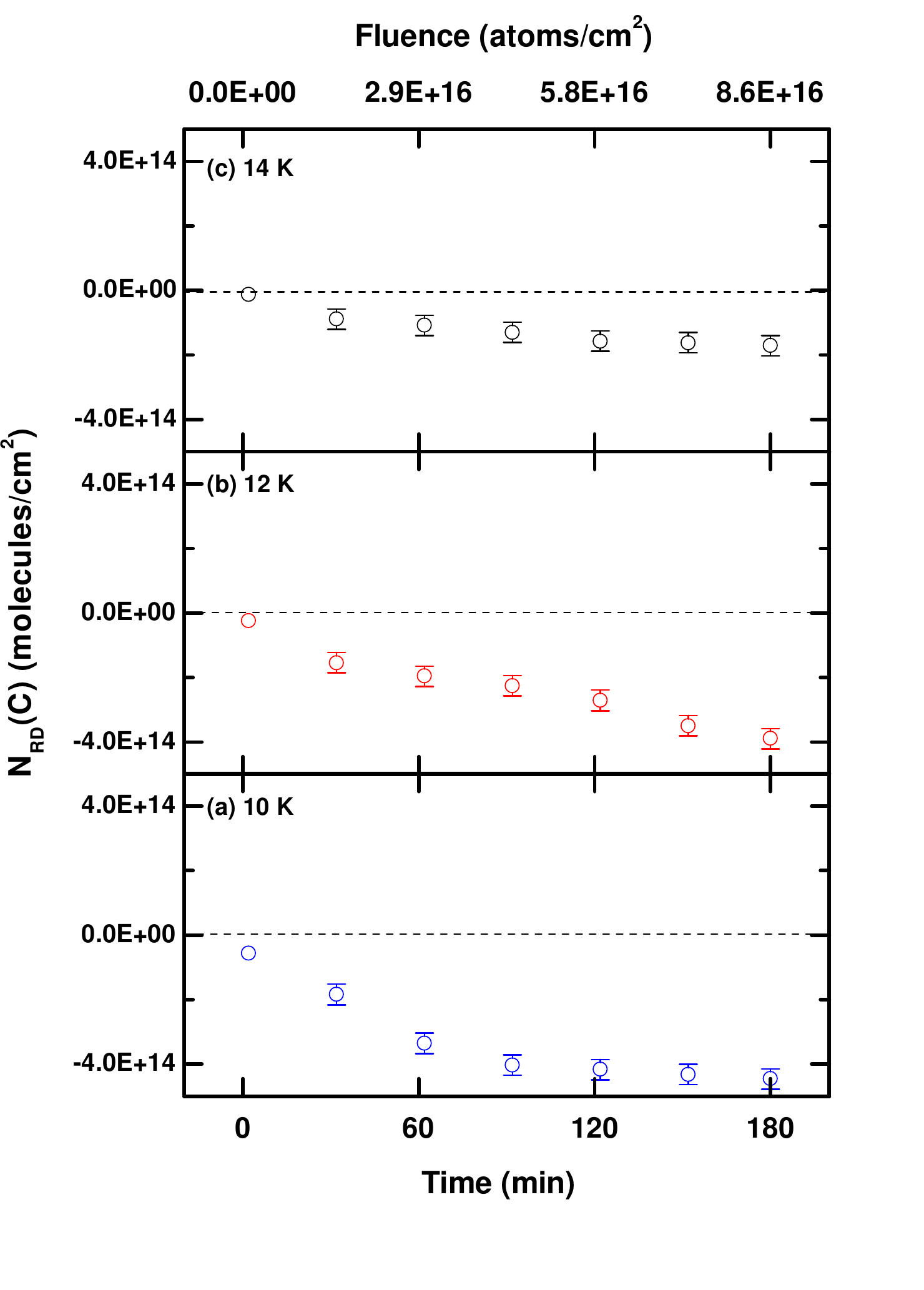}
		\caption{
			Evolution of the column density of the reactive desorption species, $\textit{N}_{RD}$(C), during a CO hydrogenation experiment with an H-atom flux of $\sim$$8$$\times$$10^{12}$ atoms cm$^{-2}$ s$^{-1}$ over $180$ minutes at (a) $10$ K, (b) $12$ K, and (c) $14$ K, respectively.
		}
		\label{Fig6.5}
	\end{center}
\end{figure}

\begin{figure}
	\begin{center}
		\includegraphics[width=98mm]{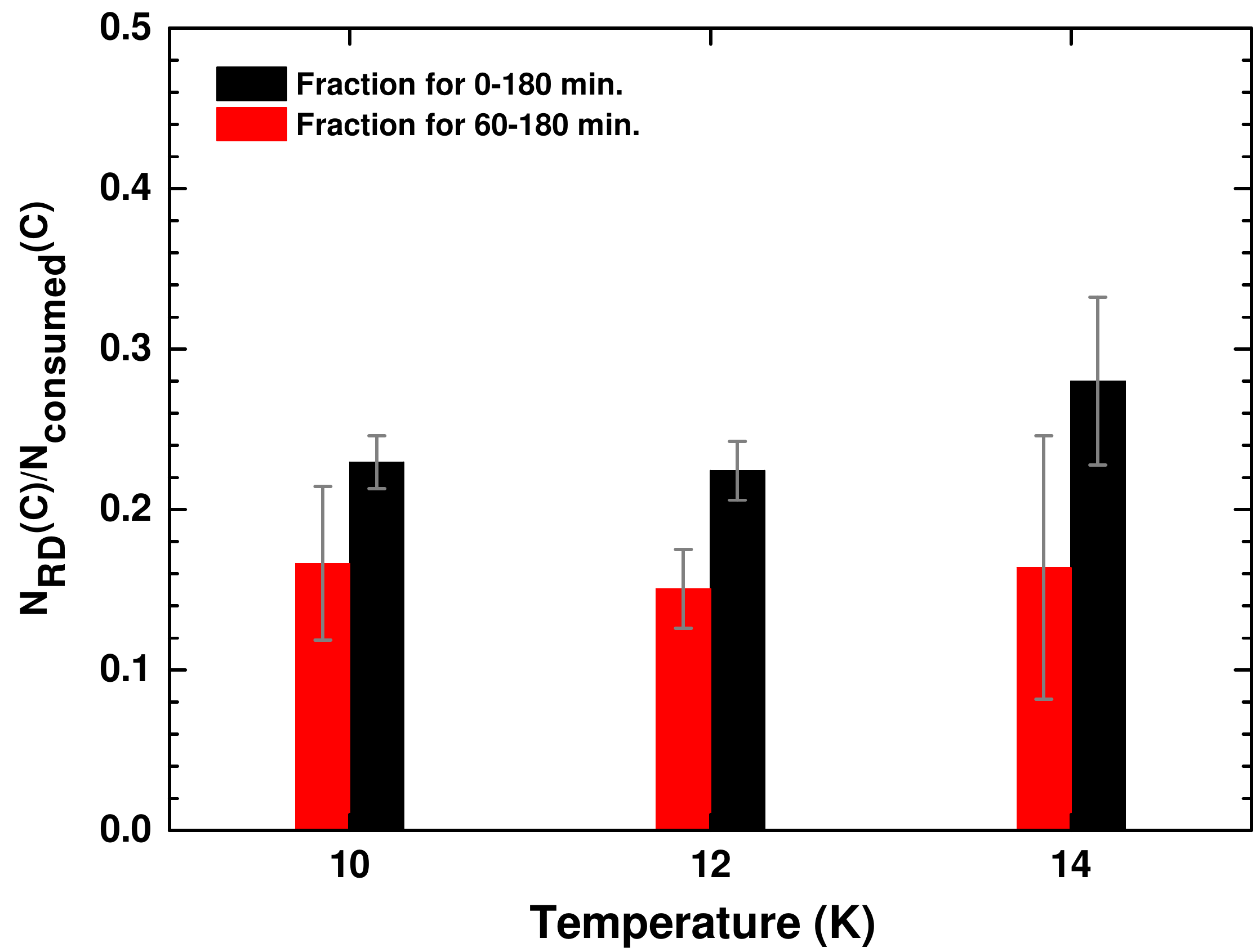}
		\caption{
			$\textit{N}_{RD}$(C)/$\textit{N}_\text{consumed}$(C) ratios obtained after the bombardment of predeposited CO ice with $\sim$$8$$\times$$10^{12}$ H-atoms cm$^{-2}$ s$^{-1}$ for the full $180$ minutes (black column) and the data recorded between $60$ and $180$ minutes (red column) at $10$, $12$, or $14$ K, respectively.
		}
		\label{Fig6.6}
	\end{center}
\end{figure}

In Figure \ref{Fig6.5}, the effective RD related column densities are shown for three different ice temperatures for $180$ minutes H-atom bombardment. The increasing loss reflects that the sum of H$_2$CO and CH$_3$OH abundances is less than the amount of consumed CO molecules. At $180$ minutes, the $\textit{N}_{RD}$ (C) value reaches $\sim$$4.5$$\times$$10^{14}$, $\sim$$3.9$$\times$$10^{14}$, and $\sim$$1.7$$\times$$10^{14}$ molecules cm$^{-2}$ for $10$ K, $12$ K, and $14$ K, respectively. As discussed in Equation (\ref{Eq6.1}), $\textit{N}_{RD}$(C) reflects all products that are not recorded by the RAIRS technique in the solid state, i.e., this variable includes both the effect of reactive desorption and other elemental carbon loss channels, such as the formation of COMs with two, three, or more carbon atoms. However, the expected final abundances of these COMs cannot account for more than a few percent of the consumed CO molecules, as discussed in \citet{Fedoseev2017}. Moreover, their CO stretching vibration modes are very similar to the CO stretching vibration modes used here for the CH$_3$OH column densities calculations; the smaller fraction of carbon atoms conserved in the formed larger COMs is actually already contributing to the CH$_3$OH column densities. Non-overlapping (and generally weaker) IR absorption features of COMs, i.e., around $860$ cm$^{-1}$ for glycolaldehyde and ethylene glycol are not visible, yielding upper limits of the order of $\sim$$5$$\times$$10^{13}$ molecules cm$^{-2}$ s$^{-1}$, i.e., $<$$3$\% of the consumed CO column density. It should be noted that previous studies on solid state COM formation by \citet{Chuang2016, Chuang2017} and \citet{Fedoseev2017} were performed in co-deposition modus that are assumed to simulate interstellar ice conditions in a more representative way.

In Figure \ref{Fig6.6}, the obtained upper limits of $\textit{N}_{RD}$(C)/$\textit{N}_\text{consumed}$(C) ratios are summarized for two different time periods: from the start to the end ($0-180$ minutes) and from $60-180$ minutes. These ratios must be strictly treated as upper limits of the reactive desorption fraction for the performed hydrogenation experiments. The abundances obtained at the end of the CO ice hydrogenation experiments after $180$ minutes of H-atom bombardment are used for the calculations (black column in Figure \ref{Fig6.6}), as here the largest amounts with reaction products are reached. This results in the best peak-to-noise ratios and therefore the lowest uncertainties. The resulting $\textit{N}_{RD}$(C)/$\textit{N}_\text{consumed}$(C) fractions are $0.23$$\pm$$0.02$, $0.22$$\pm$$0.02$ and $0.28$$\pm$$0.05$ for $10$ K, $12$ K, and $14$ K, respectively. The averaged RD fraction ($\textit{N}_{RD}$(C)/$\textit{N}_\text{consumed}$(C)$_{10-14 K}$) amounts to $0.24$$\pm$$0.02$ for overall $180$ minutes.

A detailed analysis of the data presented in Figure \ref{Fig6.5} shows that $\textit{N}_{RD}$(C) does not decrease linearly with time; roughly during the first $60$ minutes, the decrease is faster (for all three investigated temperatures). This may hint at startup effects, linked to structural weaknesses in the top surface layers because of any loosely bound CO molecules that get off much easier without the need of reactive desorption, for example, upon a collisional impact. For the period from $60$ to $180$ minutes, the $\textit{N}_{RD}$(C) decrease is much more linear. To exclude any perturbing contributions, therefore, the elemental carbon loss fraction is also calculated for the values derived between $60$ and $180$ minutes of H-atom bombardment (red column in Figure \ref{Fig6.6}) resulting in somewhat lower upper values: $0.17$$\pm$$0.05$, $0.15$$\pm$$0.02$, and $0.16$$\pm$$0.08$ for $10$ K, $12$ K, and $14$ K, respectively. The resulting averaged value ($0.16$$\pm$$0.03$) amounts to $\sim$$70$\% of the overall value. The latter value is likely more precise, as it does not include startup effects. However, to stay on the safe side, we report here the overall averaged RD fraction.

\subsection{Elemental carbon budget in H$_2$CO and CH$_3$OH hydrogenation experiments}\label{ch6Elemental Carbon Budget in H$_2$CO and CH$_3$OH Hydrogenation Experiments}

\begin{figure}
	\begin{center}
		\includegraphics[width=98mm]{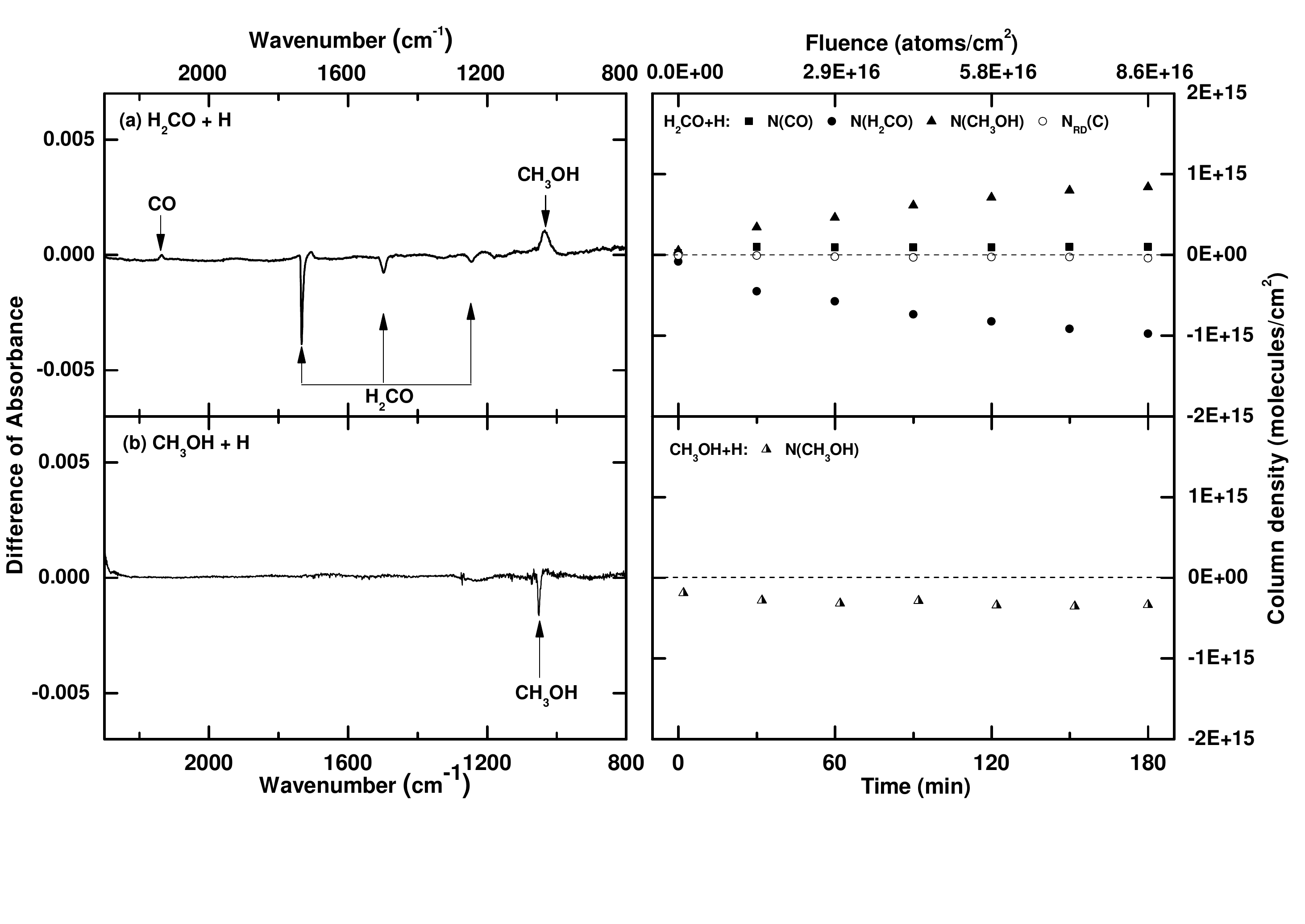}
		\caption{
			Left: RAIR difference spectra obtained after the exposure of $5$$\times$$10^{15}$ molecules cm$^{-2}$ pre-deposited (a) H$_2$CO and $8$$\times$$10^{15}$ molecules cm$^{-2}$ pre-deposited (b) CH$_3$OH to $\sim$$8$$\times$$10^{12}$ H-atoms cm$^{-2}$ s$^{-1}$ for $180$ minutes at $10$ K. Right: evolution of the surface column density of CO, H$_2$CO, and CH$_3$OH during H$_2$CO and CH$_3$OH hydrogenation over $180$ minutes at $10$ K.
		}
		\label{Fig6.7}
	\end{center}
\end{figure}

A very similar set of experiments is used to investigate reactive desorption upon H$_2$CO and CH$_3$OH hydrogenation. Figure \ref{Fig6.7} (left panel) presents the RAIR difference spectra obtained after hydrogenation of pre-deposited (a) H$_2$CO and (b) CH$_3$OH ice for an H-atom flux of $\sim$$8$$\times$$10^{12}$ atoms cm$^{-2}$ s$^{-1}$ for $180$ minutes at $10$ K. In spectrum (a), the negative peaks at $1727$, $1497$, and $1250$ cm$^{-1}$ show the depletion of the initially deposited H$_2$CO, while the positive peaks indicate the formation of newly formed products, i.e., CO ($2142$ cm$^{-1}$) and CH$_3$OH ($1030$ cm$^{-1}$). The formation of CO through H-atom abstraction reactions from H$_2$CO and of CH$_3$OH through H-atom addition reactions to H$_2$CO have been reported previously by \citet{Hidaka2004} and \citet{Chuang2016}.

In contrast, the CH$_3$OH hydrogenation experiments (Figure \ref{Fig6.7}(b)) do not result in any efficient consumption of CH$_3$OH and the formation of hydrogenation products. Only a small negative CH$_3$OH peak ($1050$ cm$^{-1}$) is observed, while no secure identification of other carbon-bearing species can be made. Moreover, the consumption peak accumulates at the very beginning of the H-atom bombardment and as discussed above; this may be due to other processes than reactive desorption. As in the CO hydrogenation experiments, there is no spectroscopic evidence for radical entrapments due to high reactivity.

The evolving column density of CO, H$_2$CO, and CH$_3$OH during H-atom exposure with time is shown in the right-hand panel of Figure \ref{Fig6.7}. The consumed $\textit{N}$(H$_2$CO) saturates at $\sim$$1$$\times$$10^{15}$ molecules cm$^{-2}$ and is possibly limited by the H-atom penetration depth as discussed before. However, $\textit{N}$(CH$_3$OH) saturates immediately, indicating the structural ice changes described before occurring on the surface comprising of CH$_3$OH molecules. As discussed for the CO hydrogenation experiments, the column density of the missing carbon during the first $60$ minutes of H-atom bombardment likely includes other unwanted desorption processes. Following the same approach as described in Section \ref{ch6Elemental Carbon Budget in H$_2$CO and CH$_3$OH Hydrogenation Experiments}, an extended calculation of the H$_2$CO hydrogenation experiment shows that the sum of the consumption of H$_2$CO and the production of CO and CH$_3$OH is very small; $\sim$$2.0$$\times$$10^{13}$ molecules cm$^{-2}$, i.e., $4$\% with respect to the CO hydrogenation experiments at $10$ K. In the CH$_3$OH hydrogenation experiments, the $\textit{N}_{RD}$(C) originates only from the consumption of $\textit{N}$(CH$_3$OH) due to a poor detection of H$_2$CO and results in $4$\% with respect to the CO hydrogenation experiments at $10$ K. These small signals with large uncertainty complicate the determination of reactive desorption fractions. Future work is needed to provide more precise RD efficiencies for H$_2$CO and CH$_3$OH.

\section{Astrophysical implication and conclusions}\label{ch6Astrophysical Implication and Conclusions}

The astronomical relevance of this work is that the RD fraction used in astrochemical models (for CO hydrogenation, but likely also more general) can be based on an experimentally determined value. The laboratory data obtained here indicate that the sum of consumption and production of carbon-bearing species along the CO hydrogenation in the solid state is not zero. The overall reactive desorption fraction $\textit{f}$$_\text{overall}$ = $\textit{N}_{RD}$(C) / $\textit{N}_\text{consumed}$ (C)$_{10-14k}$ = $0.24$$\pm$$0.02$ for CO hydrogenation into CH$_3$OH is obtained under the assumption that the missing carbon fraction is fully or largely explained by the reactive desorption of hydrogenation products into the gas phase. This value is in the best case an upper limit and the actual effect of reactive desorption may be much smaller, in line with value derived from $60-180$ minutes experiments.

The obtained upper limit is the cumulative effect of all involved reactions. This includes H-atom addition and H-atom abstraction reactions with the stable species (CO, H$_2$CO, and CH$_3$OH) as well as H-atom reactions with intermediate radicals (HCO, CH$_3$O, CH$_2$OH). Although the resulting number is an overall value, it is possible to derive the effective desorption fraction for a single H-atom reaction ($\textit{f}$$_\text{single}$) by assuming an identical efficiency for each reaction in the hydrogenation process \citep{Garrod2007}. The total fraction of the species left on the surface, i.e., ($1$ - $\textit{f}$$_\text{overall}$), is equal to the left fraction of a single reaction, i.e., ($\textit{f}$$_\text{single}$), to the power $\textit{n}$:
\begin{equation}\label{Eq6.5}
(1-\textit{f}_\text{single})^{\textit{n}}=1-\textit{f}_\text{overall}
\end{equation}
where $\textit{n}$ is the number of reactions in the hydrogenation scheme. $\textit{f}$$_\text{single}$ can be obtained from the $\textit{f}$$_\text{overall}$ measured in our experiments and depends on the average number of H-atom addition and abstraction steps occurring upon conversion of CO into CH$_3$OH (see Figure \ref{Fig6.1}). For an astrochemical simulation in which only H-atom addition reactions are taken into account, the number of steps resulting in the conversion of CO molecules into CH$_3$OH is four ($\textit{n}$ = $4$):
\begin{equation}\label{Eq6.6}
\rm CO\longrightarrow HCO\longrightarrow H_2CO\longrightarrow CH_3O\longrightarrow CH_3OH
\end{equation}
According to Equation (\ref{Eq6.5}), this gives $\textit{f}$$_\text{single}$ is $\leq$$0.066$$\pm$$0.006$. This value can be treated as the upper limit for the effective reactive desorption fraction at each of the H-atom addition steps. For a scenario in which H-atom induced abstraction reactions are also considered, this value further decreases. Assuming that both addition and abstraction reactions contributed equally efficient to the overall reactive desorption fraction along the hydrogenation sequence, e.g.,
\begin{multline}\label{Eq6.7}
\rm CO\longrightarrow HCO\longrightarrow CO\longrightarrow HCO\longrightarrow H_2CO\longrightarrow HCO\longrightarrow H_2CO\longrightarrow \\
\rm CH_3O\longrightarrow H_2CO\longrightarrow CH_3O\longrightarrow CH_3OH\longrightarrow CH_2OH\longrightarrow CH_3OH.
\end{multline}
The effective reactive desorption fraction per reaction of $\textit{f}$$_\text{single}$ is $\leq$$0.023$$\pm$$0.002$, taking $\textit{n}$ = $12$. Such effective reactive desorption fractions per hydrogenation reaction can be utilized in the models where both H-atom addition and abstraction steps are incorporated. It should be noted that at the current stage no clear understanding on the number of the involved reaction steps is achieved \citep{Nagaoka2005, Hidaka2009, Minissale2016a}. In general, a higher number of reaction steps will yield a lower effective reactive desorption fraction per single reaction. Clear data on the relative contribution of the involved H-atom addition and abstraction steps do not exist in the literature and cannot be reliably concluded from the current study.

The effective reactive desorption fractions concluded in both the aforementioned scenarios are well within the $0.01-0.10$ range of reactive desorption fractions currently used as a free parameter in astrochemical models to explain the transfer of species into the gas phase following their formation in the solid state \citep{Garrod2007, Vasyunin2013b, Balucani2015, Chang2016}. These values are clearly much lower than the $40$\%$-$$60$\% values reported earlier for other surfaces, but as discussed before, these involved experiments in the sub-monolayer regime. Values of the order of $0.01-0.1$, however, are large enough to support a solid-state mechanism enhancing abundances of small carbon-bearing species into the gas phase. This is relevant from two points of view. On one hand, it provides an intrinsic non-thermal desorption mechanism that may explain the gas-phase observation of molecules that should be fully frozen out for the low temperatures in cold clouds. \citet{Garrod2007} show that models including chemical desorption only induced by association reactions require a reactive desorption fraction of $0.03$ to optimally represent the observed gaseous CH$_3$OH abundances for L134N \citep{Dickens2000}. On the other hand, RD may offer the necessary precursors for proposed gas-phase reaction networks that result in the formation of complex molecules in addition to COMs already formed in the solid state.

The astronomical take home message is that the experimental results presented here yield an effective reactive desorption fraction for CO hydrogenation that constrain the desorption efficiency previously used in astrochemical simulations to explain astronomical observations toward pre-stellar cores. The derived values give upper limits and the real effect of RD may be smaller.

\bibliography{Ref}{}

\begin{thebibliography}{}
\expandafter\ifx\csname natexlab\endcsname\relax\def\natexlab#1{#1}\fi

\bibitem[{{Andersson} \& {van Dishoeck}(2008)}]{Andersson2008}
{Andersson}, S., \& {van Dishoeck}, E.~F. 2008, \aap, 491, 907

\bibitem[{{Arasa} {et~al.}(2015){Arasa}, {Koning}, {Kroes}, {Walsh}, \& {van
  Dishoeck}}]{Arasa2015}
{Arasa}, C., {Koning}, J., {Kroes}, G.-J., {Walsh}, C., \& {van Dishoeck},
  E.~F. 2015, \aap, 575, A121

\bibitem[{{Bacmann} {et~al.}(2012){Bacmann}, {Taquet}, {Faure}, {Kahane}, \&
  {Ceccarelli}}]{Bacmann2012}
{Bacmann}, A., {Taquet}, V., {Faure}, A., {Kahane}, C., \& {Ceccarelli}, C.
  2012, \aap, 541, L12

\bibitem[{{Balucani} {et~al.}(2015){Balucani}, {Ceccarelli}, \&
  {Taquet}}]{Balucani2015}
{Balucani}, N., {Ceccarelli}, C., \& {Taquet}, V. 2015, \mnras, 449, L16

\bibitem[{{Baratta} \& {Palumbo}(1998)}]{Baratta1998}
{Baratta}, G.~A., \& {Palumbo}, M.~E. 1998, Journal of the Optical Society of
  America A, 15, 3076

\bibitem[{{Bertin} {et~al.}(2013){Bertin}, {Fayolle}, {Romanzin}, {Poderoso},
  {Michaut}, {Philippe}, {Jeseck}, {{\"O}berg}, {Linnartz}, \&
  {Fillion}}]{Bertin2013}
{Bertin}, M., {Fayolle}, E.~C., {Romanzin}, C., {et~al.} 2013, \apj, 779, 120

\bibitem[{{Bertin} {et~al.}(2016){Bertin}, {Romanzin}, {Doronin}, {Philippe},
  {Jeseck}, {Ligterink}, {Linnartz}, {Michaut}, \& {Fillion}}]{Bertin2016}
{Bertin}, M., {Romanzin}, C., {Doronin}, M., {et~al.} 2016, \apjl, 817, L12

\bibitem[{{Boogert} {et~al.}(2015){Boogert}, {Gerakines}, \&
  {Whittet}}]{Boogert2015}
{Boogert}, A.~C.~A., {Gerakines}, P.~A., \& {Whittet}, D.~C.~B. 2015, \araa,
  53, 541

\bibitem[{{Bossa} {et~al.}(2015){Bossa}, {Mat{\'e}}, {Fransen}, {Cazaux},
  {Pilling}, {Robson Monteiro Rocha}, {Ortigoso}, \& {Linnartz}}]{Bossa2015}
{Bossa}, J.-B., {Mat{\'e}}, B., {Fransen}, C., {et~al.} 2015, \apj, 814, 47

\bibitem[{{Bouilloud} {et~al.}(2015){Bouilloud}, {Fray}, {B{\'e}nilan},
  {Cottin}, {Gazeau}, \& {Jolly}}]{Bouilloud2015}
{Bouilloud}, M., {Fray}, N., {B{\'e}nilan}, Y., {et~al.} 2015, \mnras, 451,
  2145

\bibitem[{{Brunetto} {et~al.}(2008){Brunetto}, {Caniglia}, {Baratta}, \&
  {Palumbo}}]{Brunetto2008}
{Brunetto}, R., {Caniglia}, G., {Baratta}, G.~A., \& {Palumbo}, M.~E. 2008,
  \apj, 686, 1480

\bibitem[{{Butscher} {et~al.}(2015){Butscher}, {Duvernay}, {Theule}, {Danger},
  {Carissan}, {Hagebaum-Reignier}, \& {Chiavassa}}]{Butscher2015}
{Butscher}, T., {Duvernay}, F., {Theule}, P., {et~al.} 2015, \mnras, 453, 1587

\bibitem[{{Cazaux} {et~al.}(2015){Cazaux}, {Bossa}, {Linnartz}, \&
  {Tielens}}]{Cazaux2015}
{Cazaux}, S., {Bossa}, J.-B., {Linnartz}, H., \& {Tielens}, A.~G.~G.~M. 2015,
  \aap, 573, A16

\bibitem[{{Cernicharo} {et~al.}(2012){Cernicharo}, {Marcelino}, {Roueff},
  {Gerin}, {Jim{\'e}nez-Escobar}, \& {Mu{\~n}oz Caro}}]{Cernicharo2012}
{Cernicharo}, J., {Marcelino}, N., {Roueff}, E., {et~al.} 2012, \apjl, 759, L43

\bibitem[{{Chang} \& {Herbst}(2012)}]{Chang2012}
{Chang}, Q., \& {Herbst}, E. 2012, \apj, 759, 147

\bibitem[{{Chang} \& {Herbst}(2016)}]{Chang2016}
---. 2016, \apj, 819, 145

\bibitem[{{Charnley}(1997)}]{Charnley1997}
{Charnley}, S.~B. 1997, \mnras, 291, 455

\bibitem[{{Chuang} {et~al.}(2016){Chuang}, {Fedoseev}, {Ioppolo}, {van
  Dishoeck}, \& {Linnartz}}]{Chuang2016}
{Chuang}, K.-J., {Fedoseev}, G., {Ioppolo}, S., {van Dishoeck}, E.~F., \&
  {Linnartz}, H. 2016, \mnras, 455, 1702

\bibitem[{{Chuang} {et~al.}(2017){Chuang}, {Fedoseev}, {Qasim}, {Ioppolo}, {van
  Dishoeck}, \& {Linnartz}}]{Chuang2017}
{Chuang}, K.-J., {Fedoseev}, G., {Qasim}, D., {et~al.} 2017, \mnras, 467, 2552

\bibitem[{{Cruz-Diaz} {et~al.}(2016){Cruz-Diaz}, {Mart{\'{\i}}n-Dom{\'e}nech},
  {Mu{\~n}oz Caro}, \& {Chen}}]{Cruz-Diaz2016}
{Cruz-Diaz}, G.~A., {Mart{\'{\i}}n-Dom{\'e}nech}, R., {Mu{\~n}oz Caro}, G.~M.,
  \& {Chen}, Y.-J. 2016, \aap, 592, A68

\bibitem[{{Cuppen} {et~al.}(2010){Cuppen}, {Ioppolo}, {Romanzin}, \&
  {Linnartz}}]{Cuppen2010}
{Cuppen}, H.~M., {Ioppolo}, S., {Romanzin}, C., \& {Linnartz}, H. 2010, Phys.
  Chem. Chem. Phys., 12, 12077

\bibitem[{{Cuppen} {et~al.}(2011){Cuppen}, {Penteado}, {Isokoski}, {van der
  Marel}, \& {Linnartz}}]{Cuppen2011}
{Cuppen}, H.~M., {Penteado}, E.~M., {Isokoski}, K., {van der Marel}, N., \&
  {Linnartz}, H. 2011, \mnras, 417, 2809

\bibitem[{{Cuppen} {et~al.}(2009){Cuppen}, {van Dishoeck}, {Herbst}, \&
  {Tielens}}]{Cuppen2009}
{Cuppen}, H.~M., {van Dishoeck}, E.~F., {Herbst}, E., \& {Tielens}, A.~G.~G.~M.
  2009, \aap, 508, 275

\bibitem[{{Dickens} {et~al.}(2000){Dickens}, {Irvine}, {Snell}, {Bergin},
  {Schloerb}, {Pratap}, \& {Miralles}}]{Dickens2000}
{Dickens}, J.~E., {Irvine}, W.~M., {Snell}, R.~L., {et~al.} 2000, \apj, 542,
  870

\bibitem[{{Duley} \& {Williams}(1993)}]{Duley1993}
{Duley}, W.~W., \& {Williams}, D.~A. 1993, \mnras, 260, 37

\bibitem[{{Enoch} {et~al.}(2008){Enoch}, {Evans}, {Sargent}, {Glenn},
  {Rosolowsky}, \& {Myers}}]{Enoch2008}
{Enoch}, M.~L., {Evans}, II, N.~J., {Sargent}, A.~I., {et~al.} 2008, \apj, 684,
  1240

\bibitem[{{Fedoseev} {et~al.}(2017){Fedoseev}, {Chuang}, {Ioppolo}, {Qasim},
  {van Dishoeck}, \& {Linnartz}}]{Fedoseev2017}
{Fedoseev}, G., {Chuang}, K.-J., {Ioppolo}, S., {et~al.} 2017, \apj, 842, 52

\bibitem[{{Fedoseev} {et~al.}(2016){Fedoseev}, {Chuang}, {van Dishoeck},
  {Ioppolo}, \& {Linnartz}}]{Fedoseev2016}
{Fedoseev}, G., {Chuang}, K.-J., {van Dishoeck}, E.~F., {Ioppolo}, S., \&
  {Linnartz}, H. 2016, \mnras, 460, 4297

\bibitem[{{Fedoseev} {et~al.}(2015{\natexlab{a}}){Fedoseev}, {Cuppen},
  {Ioppolo}, {Lamberts}, \& {Linnartz}}]{Fedoseev2015b}
{Fedoseev}, G., {Cuppen}, H.~M., {Ioppolo}, S., {Lamberts}, T., \& {Linnartz},
  H. 2015{\natexlab{a}}, \mnras, 448, 1288

\bibitem[{{Fedoseev} {et~al.}(2015{\natexlab{b}}){Fedoseev}, {Ioppolo}, \&
  {Linnartz}}]{Fedoseev2015a}
{Fedoseev}, G., {Ioppolo}, S., \& {Linnartz}, H. 2015{\natexlab{b}}, \mnras,
  446, 449

\bibitem[{{Fredon} {et~al.}(2017){Fredon}, {Lamberts}, \&
  {Cuppen}}]{Fredon2017}
{Fredon}, A., {Lamberts}, T., \& {Cuppen}, H.~M. 2017, \apj, 849, 125

\bibitem[{{Fuchs} {et~al.}(2009){Fuchs}, {Cuppen}, {Ioppolo}, {Romanzin},
  {Bisschop}, {Andersson}, {van Dishoeck}, \& {Linnartz}}]{Fuchs2009}
{Fuchs}, G.~W., {Cuppen}, H.~M., {Ioppolo}, S., {et~al.} 2009, \aap, 505, 629

\bibitem[{{Fulvio} {et~al.}(2009){Fulvio}, {Sivaraman}, {Baratta}, {Palumbo},
  \& {Mason}}]{Fulvio2009}
{Fulvio}, D., {Sivaraman}, B., {Baratta}, G.~A., {Palumbo}, M.~E., \& {Mason},
  N.~J. 2009, Spectrochimica Acta Part A: Molecular Spectroscopy, 72, 1007

\bibitem[{{Garrod} {et~al.}(2006){Garrod}, {Park}, {Caselli}, \&
  {Herbst}}]{Garrod2006}
{Garrod}, R., {Park}, I.~H., {Caselli}, P., \& {Herbst}, E. 2006, Faraday
  Discussions, 133, 51

\bibitem[{{Garrod} {et~al.}(2007){Garrod}, {Wakelam}, \& {Herbst}}]{Garrod2007}
{Garrod}, R.~T., {Wakelam}, V., \& {Herbst}, E. 2007, \aap, 467, 1103

\bibitem[{{Geppert} {et~al.}(2006){Geppert}, {Hamberg}, {Thomas},
  {{\"O}sterdahl}, {Hellberg}, {Zhaunerchyk}, {Ehlerding}, {Millar}, {Roberts},
  {Semaniak}, {Ugglas}, {K{\"a}llberg}, {Simonsson}, {Kaminska}, \&
  {Larsson}}]{Geppert2006}
{Geppert}, W.~D., {Hamberg}, M., {Thomas}, R.~D., {et~al.} 2006, Faraday
  Discussions, 133, 177

\bibitem[{{Hasegawa} {et~al.}(1992){Hasegawa}, {Herbst}, \&
  {Leung}}]{Hasegawa1992}
{Hasegawa}, T.~I., {Herbst}, E., \& {Leung}, C.~M. 1992, \apjs, 82, 167

\bibitem[{{Hidaka} {et~al.}(2009){Hidaka}, {Watanabe}, {Kouchi}, \&
  {Watanabe}}]{Hidaka2009}
{Hidaka}, H., {Watanabe}, M., {Kouchi}, A., \& {Watanabe}, N. 2009, \apj, 702,
  291

\bibitem[{{Hidaka} {et~al.}(2011){Hidaka}, {Watanabe}, {Kouchi}, \&
  {Watanabe}}]{Hidaka2011}
---. 2011, Phys. Chem. Chem. Phys., 13, 15798

\bibitem[{{Hidaka} {et~al.}(2004){Hidaka}, {Watanabe}, {Shiraki}, {Nagaoka}, \&
  {Kouchi}}]{Hidaka2004}
{Hidaka}, H., {Watanabe}, N., {Shiraki}, T., {Nagaoka}, A., \& {Kouchi}, A.
  2004, \apj, 614, 1124

\bibitem[{{Hiraoka} {et~al.}(1998){Hiraoka}, {Miyagoshi}, {Takayama},
  {Yamamoto}, \& {Kihara}}]{Hiraoka1998}
{Hiraoka}, K., {Miyagoshi}, T., {Takayama}, T., {Yamamoto}, K., \& {Kihara}, Y.
  1998, \apj, 498, 710

\bibitem[{{Hiraoka} {et~al.}(1994){Hiraoka}, {Ohashi}, {Kihara}, {Yamamoto},
  {Sato}, \& {Yamashita}}]{Hiraoka1994}
{Hiraoka}, K., {Ohashi}, N., {Kihara}, Y., {et~al.} 1994, Chemical Physics
  Letters, 229, 408

\bibitem[{{Hiraoka} {et~al.}(1995){Hiraoka}, {Yamashita}, {Yachi}, {Aruga},
  {Sato}, \& {Muto}}]{Hiraoka1995}
{Hiraoka}, K., {Yamashita}, A., {Yachi}, Y., {et~al.} 1995, \apj, 443, 363

\bibitem[{{Hollenberg} \& {Dows}(1961)}]{Hollenberg1961}
{Hollenberg}, J.~L., \& {Dows}, D.~A. 1961, \jcp, 34, 1061

\bibitem[{{Ioppolo} {et~al.}(2008){Ioppolo}, {Cuppen}, {Romanzin}, {van
  Dishoeck}, \& {Linnartz}}]{Ioppolo2008}
{Ioppolo}, S., {Cuppen}, H.~M., {Romanzin}, C., {van Dishoeck}, E.~F., \&
  {Linnartz}, H. 2008, \apj, 686, 1474

\bibitem[{{Ioppolo} {et~al.}(2010){Ioppolo}, {Cuppen}, {Romanzin}, {van
  Dishoeck}, \& {Linnartz}}]{Ioppolo2010}
---. 2010, Phys. Chem. Chem. Phys., 12, 12065

\bibitem[{{Ioppolo} {et~al.}(2013){Ioppolo}, {Fedoseev}, {Lamberts},
  {Romanzin}, \& {Linnartz}}]{Ioppolo2013}
{Ioppolo}, S., {Fedoseev}, G., {Lamberts}, T., {Romanzin}, C., \& {Linnartz},
  H. 2013, Review of Scientific Instruments, 84, 073112

\bibitem[{{Ivlev} {et~al.}(2015){Ivlev}, {Padovani}, {Galli}, \&
  {Caselli}}]{Ivlev2015}
{Ivlev}, A.~V., {Padovani}, M., {Galli}, D., \& {Caselli}, P. 2015, \apj, 812,
  135

\bibitem[{{Jim{\'e}nez-Serra} {et~al.}(2016){Jim{\'e}nez-Serra}, {Vasyunin},
  {Caselli}, {Marcelino}, {Billot}, {Viti}, {Testi}, {Vastel}, {Lefloch}, \&
  {Bachiller}}]{Jimenez-Serra2016}
{Jim{\'e}nez-Serra}, I., {Vasyunin}, A.~I., {Caselli}, P., {et~al.} 2016,
  \apjl, 830, L6

\bibitem[{{Kerkhof} {et~al.}(1999){Kerkhof}, {Schutte}, \&
  {Ehrenfreund}}]{Kerkhof1999}
{Kerkhof}, O., {Schutte}, W.~A., \& {Ehrenfreund}, P. 1999, \aap, 346, 990

\bibitem[{{Leger} {et~al.}(1985){Leger}, {Jura}, \& {Omont}}]{Leger1985}
{Leger}, A., {Jura}, M., \& {Omont}, A. 1985, \aap, 144, 147

\bibitem[{Ligterink {et~al.}(2018)Ligterink, Walsh, Bhuin, Vissapragada, van
  Scheltinga, \& Linnartz}]{Ligterink2018}
Ligterink, N., Walsh, C., Bhuin, R., {et~al.} 2018, Astronomy \& Astrophysics,
  612, A88

\bibitem[{Linnartz {et~al.}(2015)Linnartz, Ioppolo, \& Fedoseev}]{Linnartz2015}
Linnartz, H., Ioppolo, S., \& Fedoseev, G. 2015, International Reviews in
  Physical Chemistry, 34, 205

\bibitem[{{Minissale} {et~al.}(2016{\natexlab{a}}){Minissale}, {Congiu}, \&
  {Dulieu}}]{Minissale2016a}
{Minissale}, M., {Congiu}, E., \& {Dulieu}, F. 2016{\natexlab{a}}, \aap, 585,
  A146

\bibitem[{{Minissale} {et~al.}(2016{\natexlab{b}}){Minissale}, {Moudens},
  {Baouche}, {Chaabouni}, \& {Dulieu}}]{Minissale2016b}
{Minissale}, M., {Moudens}, A., {Baouche}, S., {Chaabouni}, H., \& {Dulieu}, F.
  2016{\natexlab{b}}, \mnras, 458, 2953

\bibitem[{{Miyauchi} {et~al.}(2008){Miyauchi}, {Hidaka}, {Chigai}, {Nagaoka},
  {Watanabe}, \& {Kouchi}}]{Miyauchi2008}
{Miyauchi}, N., {Hidaka}, H., {Chigai}, T., {et~al.} 2008, Chemical Physics
  Letters, 456, 27

\bibitem[{{Nagaoka} {et~al.}(2005){Nagaoka}, {Watanabe}, \&
  {Kouchi}}]{Nagaoka2005}
{Nagaoka}, A., {Watanabe}, N., \& {Kouchi}, A. 2005, \apjl, 624, L29

\bibitem[{{{\"O}berg} {et~al.}(2010){{\"O}berg}, {Bottinelli}, {J{\o}rgensen},
  \& {van Dishoeck}}]{Oberg2010}
{{\"O}berg}, K.~I., {Bottinelli}, S., {J{\o}rgensen}, J.~K., \& {van Dishoeck},
  E.~F. 2010, \apj, 716, 825

\bibitem[{{{\"O}berg} {et~al.}(2011){{\"O}berg}, {van der Marel}, {Kristensen},
  \& {van Dishoeck}}]{Oberg2011a}
{{\"O}berg}, K.~I., {van der Marel}, N., {Kristensen}, L.~E., \& {van
  Dishoeck}, E.~F. 2011, \apj, 740, 14

\bibitem[{{Penteado} {et~al.}(2015){Penteado}, {Boogert}, {Pontoppidan},
  {Ioppolo}, {Blake}, \& {Cuppen}}]{Penteado2015}
{Penteado}, E.~M., {Boogert}, A.~C.~A., {Pontoppidan}, K.~M., {et~al.} 2015,
  \mnras, 454, 531

\bibitem[{{Pontoppidan}(2006)}]{Pontoppidan2006}
{Pontoppidan}, K.~M. 2006, \aap, 453, L47

\bibitem[{{Rivilla} {et~al.}(2017){Rivilla}, {Beltr{\'a}n}, {Cesaroni},
  {Fontani}, {Codella}, \& {Zhang}}]{Rivilla2017}
{Rivilla}, V.~M., {Beltr{\'a}n}, M.~T., {Cesaroni}, R., {et~al.} 2017, \aap,
  598, A59

\bibitem[{{Roux} {et~al.}(1980){Roux}, {Wood}, {Smith}, \& {Plyer}}]{Roux1980}
{Roux}, J.~A., {Wood}, B.~E., {Smith}, A.~M., \& {Plyer}, R.~R. 1980, {Infrared
  optical properties of thin CO, NO, CH4, HC1, N2O, O2, AR, and air cryofilms},
  Tech. rep.

\bibitem[{{Taquet} {et~al.}(2016){Taquet}, {Wirstr{\"o}m}, \&
  {Charnley}}]{Taquet2016}
{Taquet}, V., {Wirstr{\"o}m}, E.~S., \& {Charnley}, S.~B. 2016, \apj, 821, 46

\bibitem[{{Teolis} {et~al.}(2007){Teolis}, {Loeffler}, {Raut}, {Fam{\'a}}, \&
  {Baragiola}}]{Teolis2007}
{Teolis}, B.~D., {Loeffler}, M.~J., {Raut}, U., {Fam{\'a}}, M., \& {Baragiola},
  R.~A. 2007, \icarus, 190, 274

\bibitem[{{Tielens} \& {Hagen}(1982)}]{Tielens1982}
{Tielens}, A.~G.~G.~M., \& {Hagen}, W. 1982, \aap, 114, 245

\bibitem[{{Tielens} {et~al.}(1991){Tielens}, {Tokunaga}, {Geballe}, \&
  {Baas}}]{Tielens1991}
{Tielens}, A.~G.~G.~M., {Tokunaga}, A.~T., {Geballe}, T.~R., \& {Baas}, F.
  1991, \apj, 381, 181

\bibitem[{{Tschersich}(2000)}]{Tschersich2000}
{Tschersich}, K.~G. 2000, Journal of Applied Physics, 87, 2565

\bibitem[{{van Dishoeck} {et~al.}(2013){van Dishoeck}, {Herbst}, \&
  {Neufeld}}]{vanDishoeck2013}
{van Dishoeck}, E.~F., {Herbst}, E., \& {Neufeld}, D.~A. 2013, Chemical
  Reviews, 113, 9043

\bibitem[{{Vasyunin} {et~al.}(2017){Vasyunin}, {Caselli}, {Dulieu}, \&
  {Jim{\'e}nez-Serra}}]{Vasyunin2017}
{Vasyunin}, A.~I., {Caselli}, P., {Dulieu}, F., \& {Jim{\'e}nez-Serra}, I.
  2017, \apj, 842, 33

\bibitem[{{Vasyunin} \& {Herbst}(2013)}]{Vasyunin2013b}
{Vasyunin}, A.~I., \& {Herbst}, E. 2013, \apj, 769, 34

\bibitem[{{Watanabe} \& {Kouchi}(2002)}]{Watanabe2002}
{Watanabe}, N., \& {Kouchi}, A. 2002, \apjl, 571, L173

\bibitem[{{Watanabe} {et~al.}(2004){Watanabe}, {Nagaoka}, {Shiraki}, \&
  {Kouchi}}]{Watanabe2004}
{Watanabe}, N., {Nagaoka}, A., {Shiraki}, T., \& {Kouchi}, A. 2004, \apj, 616,
  638

\bibitem[{{Weast} \& {Astle}(1985)}]{Weast1985}
{Weast}, R.~C., \& {Astle}, M.~J. 1985, CRC Handbook of Data on Organic
  Compounds (Boca Raton, FL: CRC Press), 968

\bibitem[{{Westley} {et~al.}(1998){Westley}, {Baratta}, \&
  {Baragiola}}]{Westley1998}
{Westley}, M.~S., {Baratta}, G.~A., \& {Baragiola}, R.~A. 1998, \jcp, 108, 3321

\bibitem[{{Willacy} \& {Millar}(1998)}]{Willacy1998}
{Willacy}, K., \& {Millar}, T.~J. 1998, \mnras, 298, 562

\bibitem[{{Woods} {et~al.}(2012){Woods}, {Kelly}, {Viti}, {Slater}, {Brown},
  {Puletti}, {Burke}, \& {Raza}}]{Woods2012}
{Woods}, P.~M., {Kelly}, G., {Viti}, S., {et~al.} 2012, \apj, 750, 19

\bibitem[{{Zhitnikov} \& {Dmitriev}(2002)}]{Zhitnikov2002}
{Zhitnikov}, R.~A., \& {Dmitriev}, Y.~A. 2002, \aap, 386, 1129

\end{thebibliography}

\end{document}